# Bilayer magnetic composite systems made of random anisotropy hexaferrite and soft metallic magnets for stealth technology

Jaume Calvo-de la Rosa[a,b*], Jesús López-Sánchez[c,d], Joan Manel Hernàndez[a,b], Pilar Marín[c,d], and Javier Tejada[a]

[a] *Departament de Física de la Matèria Condensada, Universitat de Barcelona, Martí i Franquès 1, 08028 Barcelona, Spain*

[b] *Institut de Nanociència i Nanotecnologia (IN2UB), Universitat de Barcelona,*

*08028 Barcelona, Spain*

[c] *Instituto de Magnetismo Aplicado (IMA-UCM-ADIF), 28230 Madrid, Spain*

[d] *Departamento de Física de Materiales, Facultad de Físicas, Universidad Complutense de Madrid (UCM), 28040 Madrid, Spain*

* *Corresponding author: jaumecalvo@ub.edu*

## ABSTRACT

An experimental strong increase of the reflection loss (from 25 up to 35 dBs) and an extension of the absorption bandwidth up to 20% is measured in a set of novel functional bilayer systems. We focus our work on studying the samples in an anechoic chamber under far-field real radar conditions. Each layer consists of a composite material, typically a dielectric matrix filled with random anisotropy hexaferrite and soft metallic materials (powder or wires). Combining the two types of materials into a submillimetric bilayer structure has shown unprecedented improvements in microwave absorption capacities compared to the former absorption of each layer. The capacity to improve the shielding behavior is strongly related to each layer's permittivity, permeability, and thickness leading, therefore, to a strong control over the design of novel materials for stealth applications.

## 1. Introduction

Radio detection and ranging (radar) technology has been extensively used during the last century. Nonetheless, this technology has notably evolved, and it is still attracting much attention in fields like aviation [1,2], marine navigation [3,4] or space exploration [5], for instance. Many scientific and engineering efforts are paid to the generation, control, and detection of microwaves to improve the capacities of this technology. However, on the other side, stealth technology fights for the opposite goal: making objects invisible to radar detection [6]. Metamaterials or electromagnetic wave absorbers (EMWAs) are two different approaches to designing materials that prevent the incident waves from being reflected from the object. EMWAs may also be used for other purposes, like protecting critical systems from foreign electromagnetic (EM) radiation, such as strategic computing, control, or telecommunication structures. The appearance of electromagnetic pulse (EMP) weapons [7,8] makes EM protection a more sensitive issue day by day.

A large variety of materials have been studied as EMWAs [9], ranging from ceramics, metals, or even polymers. In this scenario, carbon-based structures [10–15] are a class of materials that are gaining momentum due to their low density, strong dielectric loss, and good mechanical properties [16,17]. Nonetheless, they are somehow limited by the complexity of their preparation and scalability. On the other hand, ferrites have shown excellent shielding capacities while keeping a good balance between the above-mentioned properties. More particularly, hexaferrites (typically $BaFe_{12}O_{19}$ or $SnFe_{12}O_{19}$) have shown particularly good microwave attenuation capacities [18,19]. Finally, soft magnetic metals and alloys have also attracted attention due to their high saturation magnetization and magnetic susceptibility. Amorphous magnetic microwires have extensively been reported as excellent microwave absorbers, due to the strong attenuation they show and their capacity to be modified for different requirements [20–22]. Their absorption frequency has been controlled through the wire length [20]. These materials were patented [23] and industrially exploited for electromagnetic shielding applications for years.

Recent theoretical studies [24–26] postulate random anisotropy magnets as excellent candidates for microwave absorption. In the same direction, recent experiments have verified that barium hexaferrite-based nanocomposites behave as random-anisotropy magnets [27], converting them into excellent candidates as broadband microwave absorbers. These nanocomposites are understood as a collection of nanocrystallites, of similar size to the domain wall width, each of them with a randomly oriented anisotropy axis. Both anisotropy and exchange interactions govern the magnetization dynamics; however, under these conditions, anisotropy becomes dominant over exchange interaction. The conclusions reported in [27] state that these ceramic random-anisotropy magnets show improved broadband absorption capacities in thin (sub-millimetric) layer systems.

The present work provides experimental evidence of a significant increase in the radar absorption capabilities of a well-known type of magnetic microwire sheets by combining them with an additional composite layer. This secondary layer comprises a dielectric matrix and a magnetic powder filler, either metallic or hexaferrite-based nanocomposites. The data reported here also provides a deep analysis of the effect that the creation of the bilayer configuration has on the absorption bandwidth. Overall, the novel composite and multilayered materials presented in this work demonstrate an unprecedented ability to enhance both absorption bandwidth and power simultaneously. That is, this research paves the way for the development of a new class of bilayers with highly competitive radar absorption capabilities.

## 2. Materials and methods

### 2.1. Materials and synthesis

In this work, we make use of multiple magnetic materials, either powder or wire-like. All the components that constitute our composites were studied and reported in previous works, which we refer to, and where all their chemical, structural, and magneto-dielectric characteristics may be found. On one hand, we used own-synthesized barium hexaferrite (BaHF) and hexaferrite-based nanocomposites, which have been proven to behave as random anisotropy magnets [27]. In that work [27] we report precise structural and magnetic data that justifies their random anisotropy behavior, as well as the complex permittivity and permeability in the microwave frequency range. These pure and composite hexaferrite nanocompounds were synthesized using conventional co-precipitation methods [28]. Stoichiometric amounts of the necessary metal salt precursors [$Ba(NO_3)_2$, $Fe(NO_3)_3 \cdot 9H_2O$, $Cu(NO_3)_2 \cdot 3H_2O$ and $Mn(NO_3)_2 \cdot 4H_2O$, all from Sigma Aldrich as received] were dissolved in distilled water and stirred for two hours. Once the reactants were dissolved, a 2M NaOH (Sigma Aldrich, as received) solution was incorporated into the solution dropwise until a basic pH of ~10. In these basic conditions, precipitation occurs. The solution was stirred while heated at 80°C for two additional hours to facilitate the completion of the reaction. The obtained suspension was subjected to 4 centrifugation cycles of 10 minutes at 3000 rpm. Between each cycle, the remaining solid product was washed with a mixture of water and ethanol at 50% in volume to remove the remaining undesired products from the previous steps. The obtained solid was then dried for 24 hours at 80°C and ground with a mortar before a final calcination process of 1 hour at 900°C, which activates the solid-state diffusion to form the desired ferrite. The ferrite is finely ground again to obtain a fine and homogenous powder.

On the other hand, metallic soft magnets are also considered in this study. Pure iron and a type of iron-based soft magnetic composite powder samples were provided by the enterprise AMES S.L. The first type of material (which we will refer to as soft magnetic material, SMM, according to literature) consists of pure iron particles with an average diameter of 100 µm. This powdered material stands out for its high magnetic permeability. The other material (referred to as soft magnetic composite, SMC) consists of 200 µm particles made of an iron core and an isolating phosphate-based isolating shell. This isolating shell prevents percolation between particles and the subsequent creation of eddy currents, which lead to power losses. This material can maintain a high permeability upon higher frequencies. The complete chemical, structural, and magnetic characteristics of these two materials may be found at [29].

The last class of materials used are magnetic microwires, which consist of an iron core wire surrounded by a glass coating. These microwires were prepared by the Taylor technique [30]. The inner core is made of $Fe_{89}BSi_3Mn_4$ and has a diameter of 12 µm. The total diameter is 20 µm, while the wires' length is 1 mm.

Table I below summarizes and identifies all these materials, which are used as fillers in our composites:

**Table I.** IDs and descriptions of the different materials used in this work. The table also includes the references where a full characterization of each component may be found.

| ID | Material | Former references |
|---|---|---|
| MW | Amorphous magnetic microwires | [20–23] |
| HF | Own synthesized BaHF | [27] |
| HF-Cu | Own synthesized BaHF + Cu nanocomposite | [27] |
| HF-Mn | Own synthesized BaHF + Mn nanocomposite | [27] |
| SMM | Pure iron particles | [29] |
| SMC | Iron particles with a phosphate-isolating shell | [29] |

In our samples, paint served as a dielectric isolating matrix in which the magnetic fillers were dispersed. Titan's Unilak white water-based enamel was used to disperse the powder samples, while Hempel's primer undercoat was employed for the microwires.

### 2.2. Sheets preparation

The magnetic powders and wires were mixed with the appropriate amount of paint and applied onto 25 cm × 25 cm polyester sheets to create layer-type samples with micrometric thickness. The required amount of filler was weighed using an analytical balance and dispersed into 50 g of liquid paint. The mixture was stirred for one hour to achieve a homogeneous and stable dispersion, then uniformly applied onto the polyester sheet. The paint was allowed to dry for 24 hours. The thickness was controlled by the number of coating cycles applied, with each cycle typically producing a dry layer of approximately 200 µm. In this work, we aim to analyze the effect that multiple design factors (such as the chemical nature of the filler, and the subsequent magnetic and dielectric properties) have on their radar absorption capacities. Table II lists the characteristics of the collection of sheets prepared for this work. In addition to the filler's chemical nature, we have mostly focused our efforts on inspecting the effect of another crucial geometrical aspect: the layer's thickness. As it may be seen, for each type of filler we have prepared three sheets with different thicknesses, to which we refer as small (S), medium (M), and large (L). All 16 fabricated samples and their main characteristics are listed in Table II. Given that the weight filling factor ($ff_W$) is constant, the sample identifiers follow the next structure: *filler_thickness*, being the thickness represented by its class type (S, M, or L).

**Table II.** Characteristics of the composite layer samples.

| # | ID | Matrix | Filler | $ff_w$ (%) | Thickness (mm) |
|---|---|---|---|---|---|
| 1 | MW | Hempel primer undercoat | MW | 1.2 | 0.280 |
| 2 | HF_S | Titan's Unilak paint | HF | 4.0 | 0.290 |
| 3 | HF_M | Titan's Unilak paint | HF | 4.0 | 0.580 |
| 4 | HF_L | Titan's Unilak paint | HF | 4.0 | 0.720 |
| 5 | HF_Cu_S | Titan's Unilak paint | HF_Cu | 4.0 | 0.340 |
| 6 | HF_Cu_M | Titan's Unilak paint | HF_Cu | 4.0 | 0.560 |
| 7 | HF_Cu_L | Titan's Unilak paint | HF_Cu | 4.0 | 0.820 |
| 8 | HF_Mn_S | Titan's Unilak paint | HF_Mn | 4.0 | 0.290 |
| 9 | HF_Mn_M | Titan's Unilak paint | HF_Mn | 4.0 | 0.540 |
| 10 | HF_Mn_L | Titan's Unilak paint | HF_Mn | 4.0 | 0.740 |
| 11 | SMM_S | Titan's Unilak paint | SMM | 4.0 | 0.320 |
| 12 | SMM_M | Titan's Unilak paint | SMM | 4.0 | 0.480 |
| 13 | SMM_L | Titan's Unilak paint | SMM | 4.0 | 0.770 |
| 14 | SMC_S | Titan's Unilak paint | SMC | 4.0 | 0.380 |
| 15 | SMC_M | Titan's Unilak paint | SMC | 4.0 | 0.500 |
| 16 | SMC_L | Titan's Unilak paint | SMC | 4.0 | 0.560 |

### 2.3. Electromagnetic characterization

The reflection loss ($R_L$) of all the samples was measured inside an anechoic chamber, which is equipped with broadband electromagnetic radiation absorbers (EMC-24PCL, ETS-Lindgren) and plates to prevent edge effects. The absorbing medium consists of a pyramidal polyurethane foam, designed with a specific geometry to minimize losses and attenuation of the incident wave, as well as to reduce noise that could affect measurements. Samples were subjected to a metallic plate to ensure total reflection. The signal was emitted by one antenna, reflected by the metallic plate behind the sample, and detected back by the other antenna ($S_{21}$ measurement). The EMCO 3160-07 horn antennas, positioned under far-field conditions relative to the optimized absorbing paint, were connected to an Agilent E8362B PNA Series Network Analyzer outside the anechoic chamber to generate and detect the reflected signals. Under these conditions, the transmission of electromagnetic waves can be considered as plane waves. 1601 points were measured between 0.5 and 18 GHz. Therefore, we work on the microwave and radar regime. The measurement system is schematized in Figure 1.

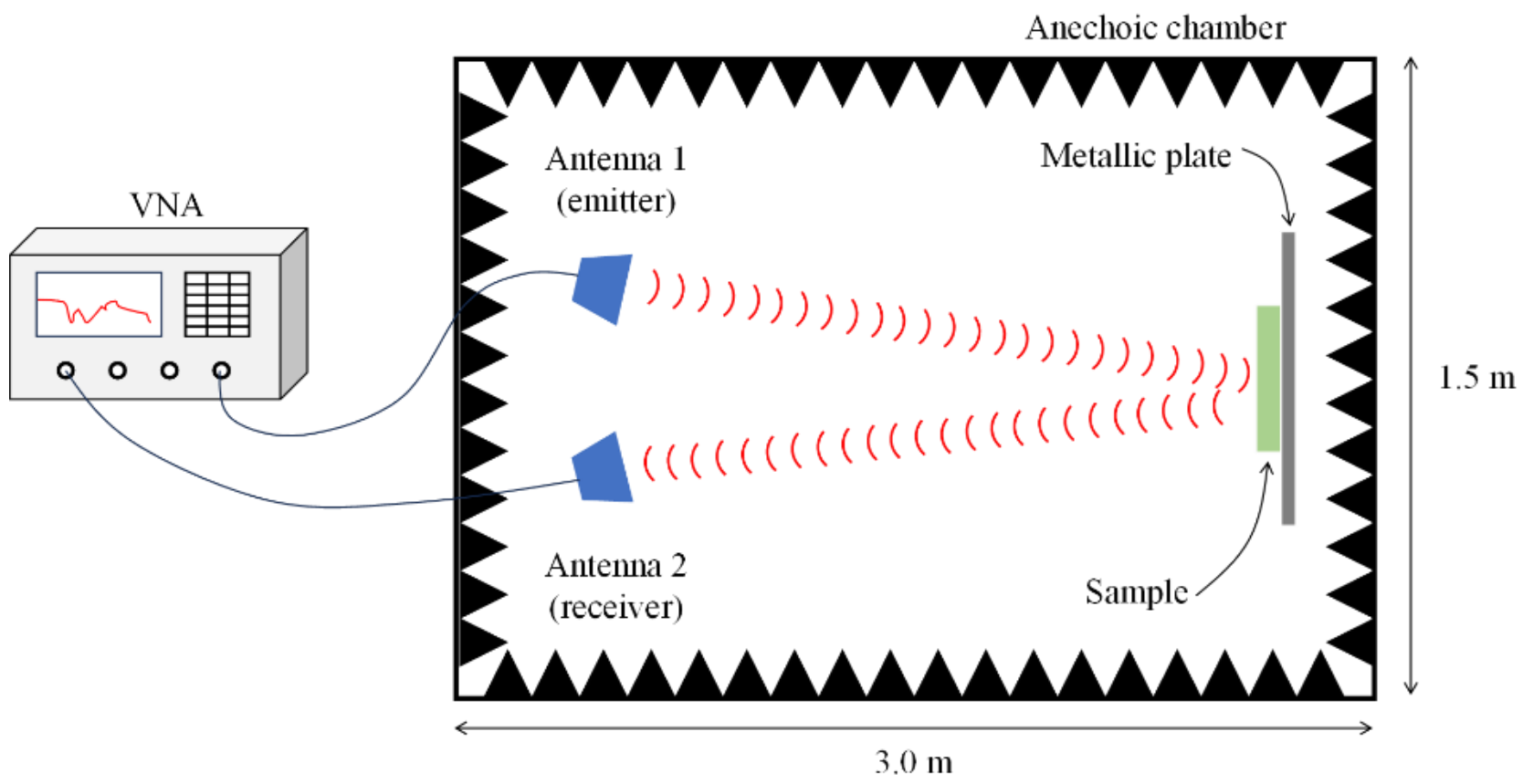

**Figure 1.** Schematic representation of the far field radar measurements done inside the anechoic chamber.

## 3. Results and discussion

We start by measuring, individually, the sheets listed in Table II. Figure 2 presents, as an example, the data for the MW sample along with the thinnest sheet from each composition. The MW sample exhibits an absorption of 25 dB at approximately 9.5 GHz, which is consistent with previous reports on the same material. [31]. On the other hand, none of the sheets containing hexaferrite or metallic powder show significant absorption in this frequency range. The same full reflection behavior was observed for both the M and L sheets as well. With these results, we conclude, without doubt, that the sheets containing hexaferrite or metallic powder are non-absorbent in this frequency range. This does not imply that these materials cannot act as radar absorbers, but rather that they do not absorb within this frequency range for the specific characteristics considered (i.e., composite layers with $ff_W$ = 4% and thicknesses between approximately 0.25 and 0.75 mm). Indeed, pure hexaferrite-based samples have been recently proven to absorb in this frequency range [27].

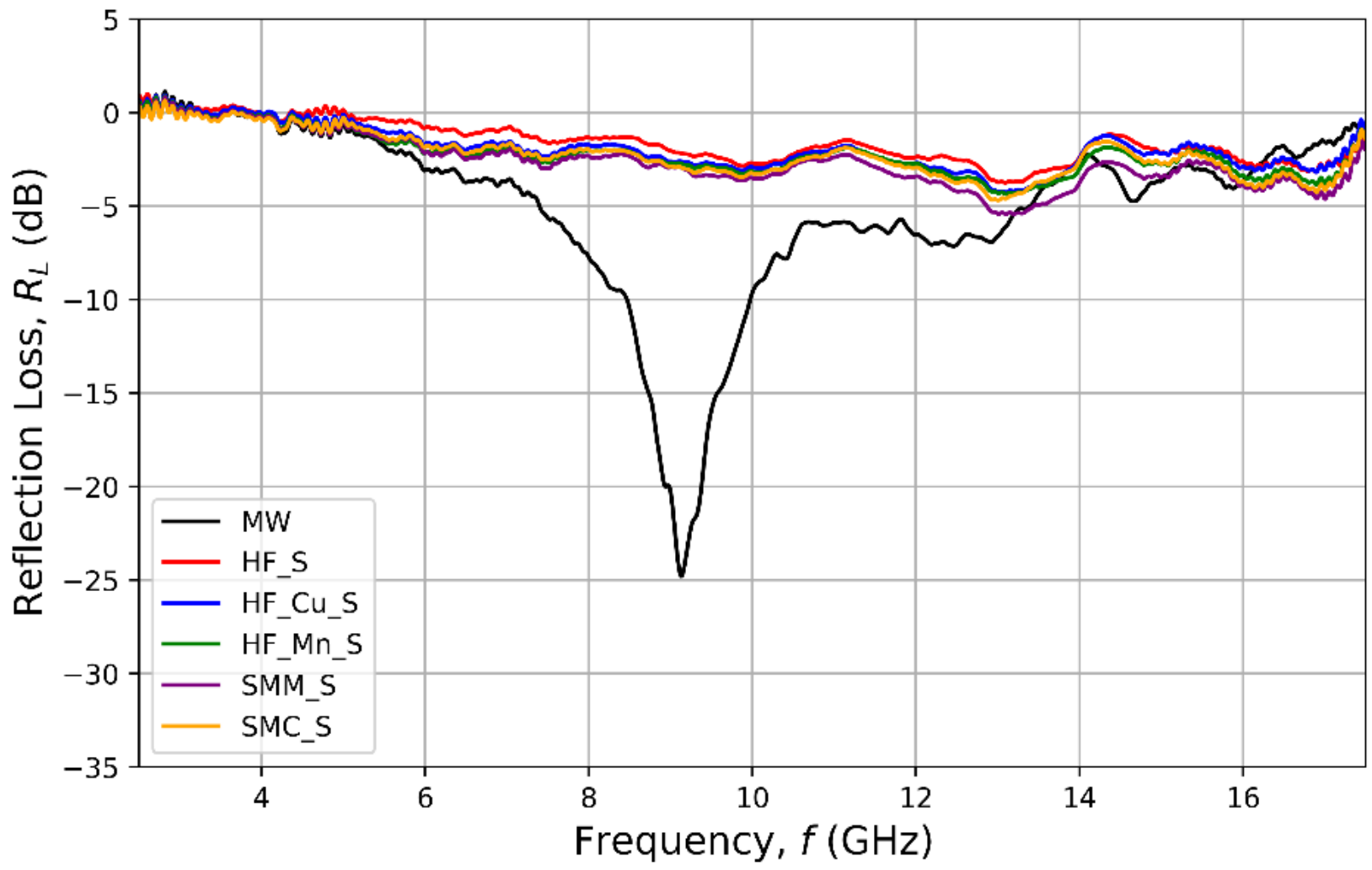

**Figure 2.** Experimental $R_L$ spectra, measured inside an anechoic chamber, for sample MW (black), HF_S (red), HF_Cu_S (blue), HF_Mn_S (green), SMM_S (purple), and SMC_S (orange).

Once the response for each sheet is clear, we start combining our reference MW sample with an additional ceramic or metal-based sheet. To begin, Figure 3 shows the obtained results when the MW reference sample is combined with an additional layer filled with HF, either pure or modified. The full spectrum is represented in the main left column, while a closer view of the peaks is provided in the supporting right-side column.

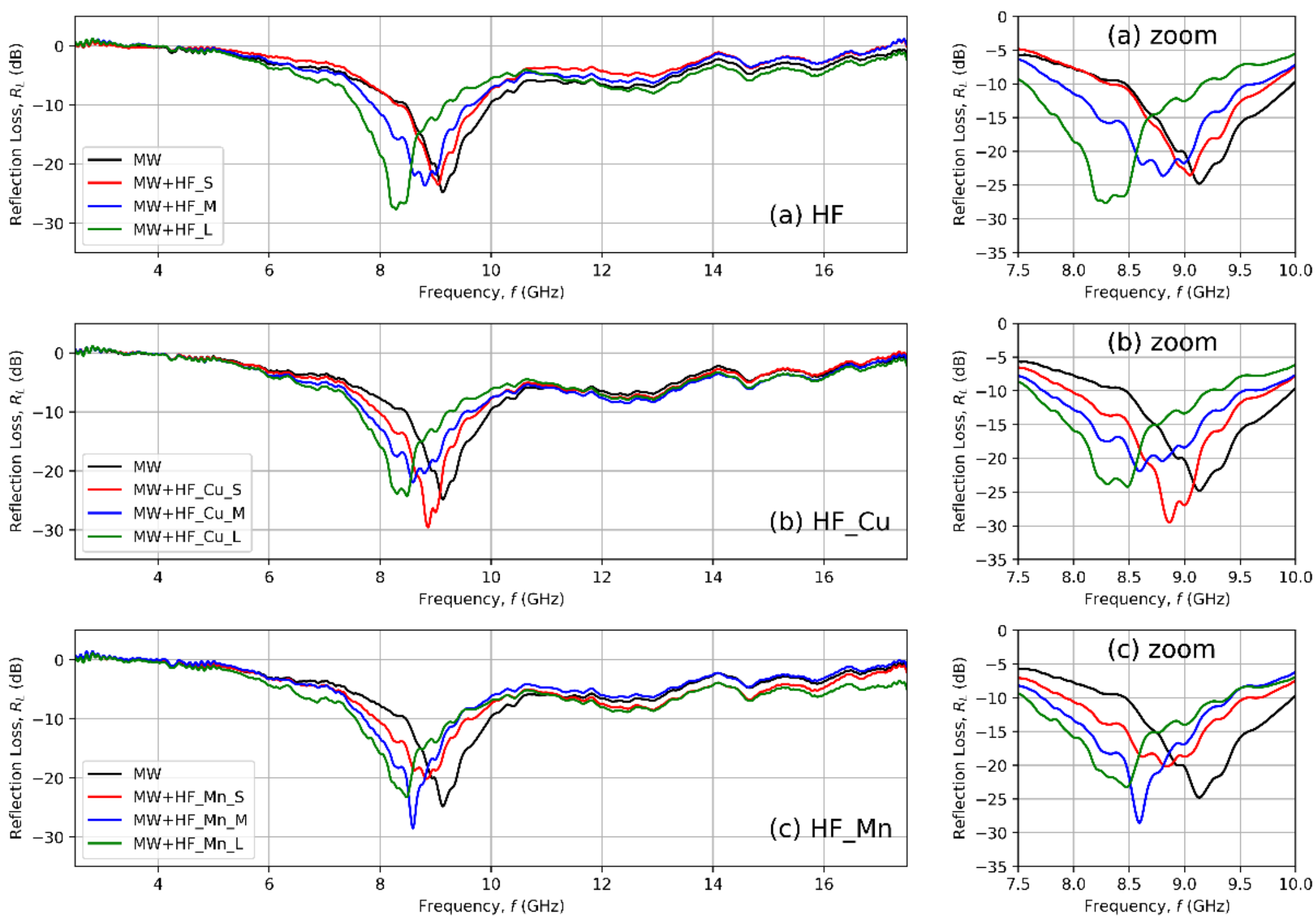

**Figure 3.** Experimental $R_L$ spectra, measured inside an anechoic chamber, for the addition of S (red), M (blue), and L (green) size distribution of (a) pure HF, (b) HF_Cu, and (c) HF_Mn powder. Left-side columns show the full spectrum between 2.5 and 17.5 GHz, while the right-side column provides an amplification close to the peaks.

In Figure 3(a) we see how an additional HF-based layer behaves. It is observed that thin (S) layers do not significantly affect absorption, whereas thicker sheets are capable of altering the response. Only for HF_L, we detect an improvement of the former MW absorption, moving $R_L$ from -25 dB to -27 dB. Additionally, a consistent shift in the absorption frequency (i.e., the position of the absorption peak) is observed as a function of the second layer thickness. As expected, the greater the sheet thickness, the lower the absorption frequency. The combination of MW+HF_L also seems to lead to a wider absorption than the former MW.

The effect is much more pronounced when HF_Cu is used instead of HF, as Figure 3(b) demonstrates. Even with the addition of HF_Cu_S, the $R_L$ increases up to -30 dB. Contrary to the previous case, thicker layers do not improve the results. This is because of the different electromagnetic properties (dielectric permittivity and magnetic permeability) of the HF_Cu compared to HF, which requires a different optimum thickness to maximize absorption. On the other hand, the frequency shift as a function of thickness is consistent with the previous results.

Finally, moving to Figure 3(c), we see that HF_Mn has an intermediate effect. In this case, the addition of a thin HF-Mn layer (S) reduces the original absorption, but it quickly increases when the intermediate sheet (M) is used. Then, it goes down again when the thickest (L) sheet is employed. A much clearer analysis of these behaviors is provided in Figure 4, where three different absorption characteristics are systematically compared as a function of the second layer thickness.

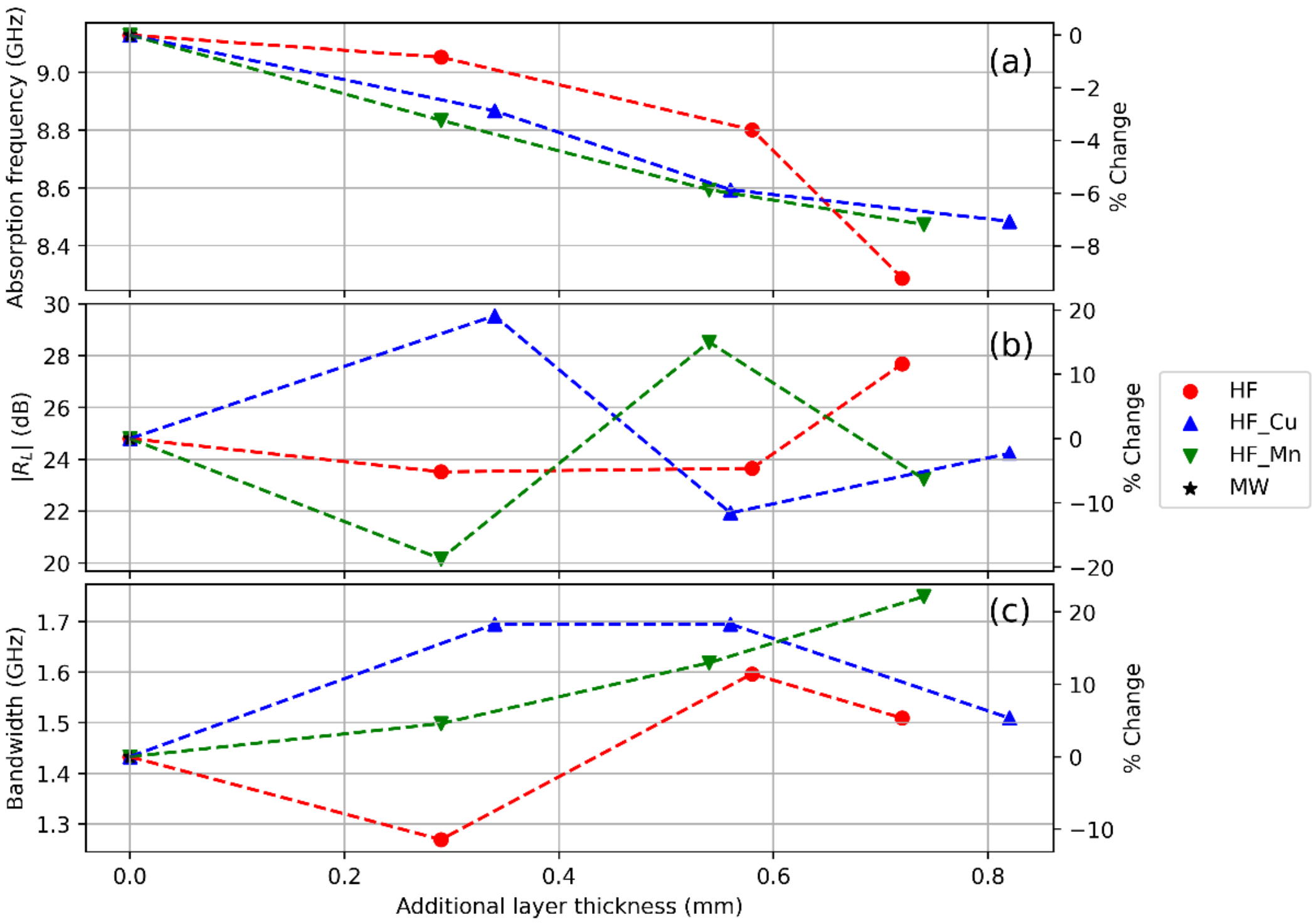

**Figure 4.** Dependence of the (a) absorption frequency, (b) maximum absolute absorption, and (c) absorption bandwidth at -10 dB as a function of the thickness of the additional layer placed on top of the MW one (black star, at the origin) and its chemical composition, being HF (red circles), HF_Cu (blue up triangles) or HF_Mn (down triangles).

In Figure 4(a), one may observe that the addition of the second layer displaces the absorption frequency to the lower region of the frequency domain. This is coherent given that the total thickness of the system is increased. Overall, with thin samples (< 1 mm) we can change the absorption frequency by 1 GHz in a continuous and controlled manner.

Figure 4(b) shows the change in the absolute powder absorption. At first look, one might think there is no solid dependence on the second layer thickness. However, there are important points that must be mentioned. First, we can see that the maximum absorption happens only at a specific thickness. If the thickness is increased or reduced, the absorption falls. Improvements of almost 30% (in dB scale) compared to the former MW absorption are achieved at specific conditions. This is in excellent agreement with the bilayer theory [32,33] which states that there is no linear dependence between the absorption and the thickness of the layers. This model, mathematically described by equations (1) and (2), defines the dependence of the $R_L$ on each layer's permittivity, permeability, and thickness. The thickness must be optimized, according to the medium's electromagnetic properties, to achieve maximum absorption. Therefore, these results could even be improved with a proper design of the first layer's thickness.

$$Z = Z_0 \frac{\sqrt{\frac{\mu_1}{\varepsilon_1}} tanh\left[\left(j\frac{2\pi f d_1}{c}\right)\sqrt{\mu_1\varepsilon_1}\right] + \sqrt{\frac{\mu_2}{\varepsilon_2}} tanh\left[\left(j\frac{2\pi f d_2}{c}\right)\sqrt{\mu_2\varepsilon_2}\right]}{1 + \sqrt{\frac{\mu_1\varepsilon_2}{\mu_2\varepsilon_1}} tanh\left[\left(j\frac{2\pi f d_1}{c}\right)\sqrt{\mu_1\varepsilon_1}\right] tanh\left[\left(j\frac{2\pi f d_2}{c}\right)\sqrt{\mu_2\varepsilon_2}\right]} \quad (1)$$

$$R_L(dB) = -20\, log \left|\frac{Z/Z_0 - 1}{Z/Z_0 + 1}\right| \quad (2)$$

In good agreement with the observations made for Figure 3, HF_Cu is the filler that increases the absorption earlier, with thinner added layers. Then, HF_Mn maximizes absorption for the intermediate (M) thickness layer, while HF is the one that requires thicker configurations to enhance the former absorption. These geometrical dependencies are in excellent agreement with [27], where random-anisotropy magnets are proved to be ideal candidates as broadband absorbers as thin coatings. This is of great importance for lightweight radar absorption applications.

All these composites possess both magnetic and dielectric components. When an electromagnetic wave interacts with such materials, it induces magnetic and dielectric polarizations, which are characterized by their complex permeability and permittivity, respectively. Recent studies [11,34,35]suggest that the increase in interfaces within these composite structures enhances interfacial polarization effects, thereby improving wave absorption. However, it is the complex impedance that governs the overall interaction between these components. The aforementioned references examining absorption in bilayer systems [32,33] highlight the necessity of achieving impedance matching between different media (e.g., air–first layer and first layer–second layer) to optimize absorption. According to the corresponding models, merely increasing permittivity, permeability, or layer thickness does not guarantee an improvement in reflection loss ($R_L$). Instead, maximizing absorption requires a precise balance between the intrinsic properties and geometrical characteristics of the bilayer system.

Finally, Figure 4(c) probably provides two of the most significant observations: (i) the addition of the second layer always expands the absorption bandwidth, and (ii) the random-anisotropy magnetic nanocomposites (HF_Cu and HF_Mn) have a better performance compared to the pure HF. These peak-broadening results also support the argument that identifies random-anisotropy magnets as ideal broadband microwave absorbers. Our experimental results prove an increase in the absorption bandwidth by up to 20% compared to the single MW layer. Again, we can see that there is an optimum thickness that maximizes the response, which depends on the specific electromagnetic properties of each sheet.

We now turn to studying the effect of replacing the ceramic-based secondary layer with a metal-based one. In this case, we keep the MW sheet as the primary reference layer and we incorporate a secondary layer on top with a metallic soft magnet dispersed on it, either SMM or SMC. The results are depicted in Figure 5.

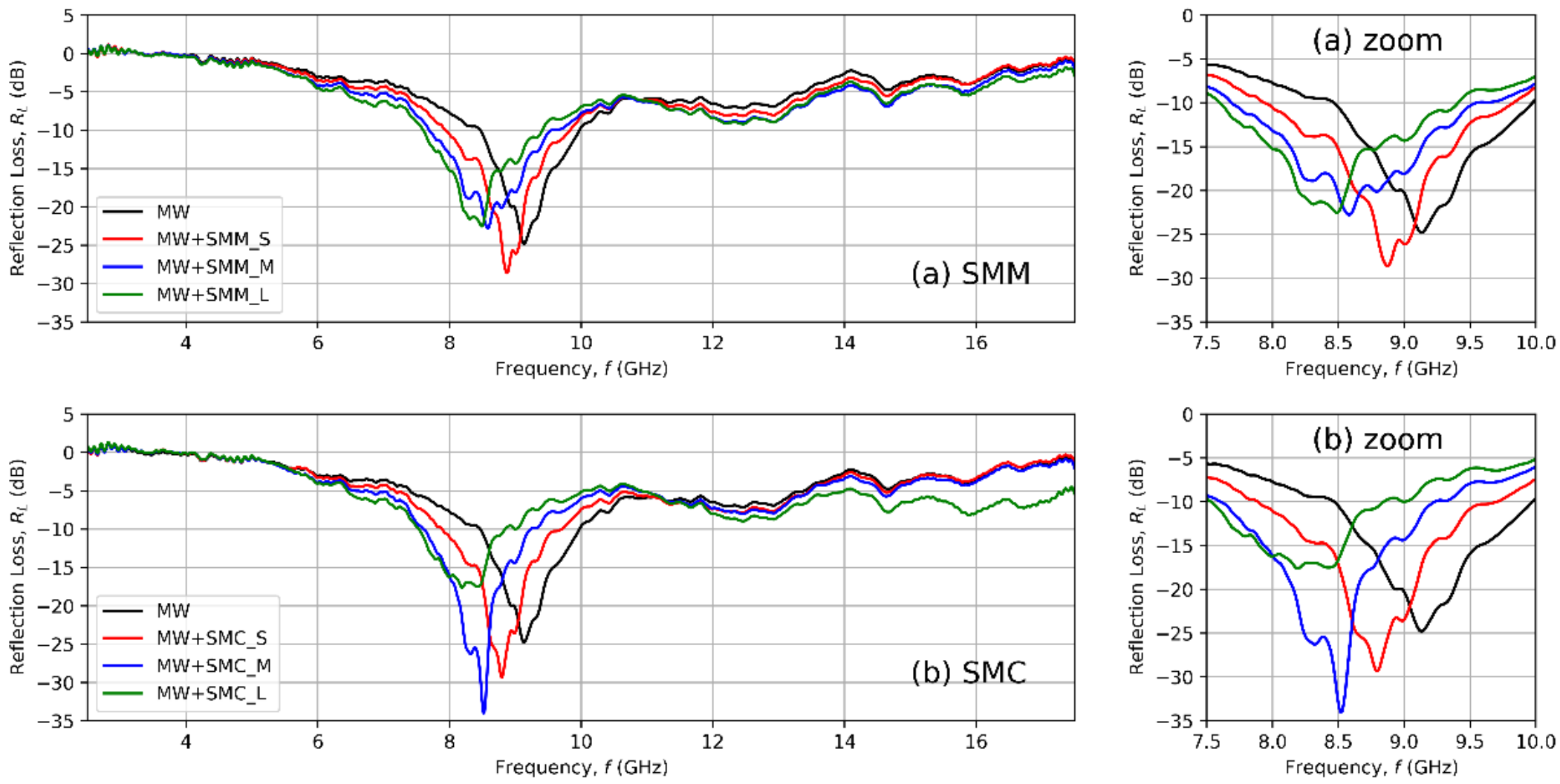

**Figure 5.** Experimental $R_L$ spectra, measured inside an anechoic chamber, for the addition of S (red), M (blue), and L (green) size distribution of (a) SMM and (b) SMC powder. Left-side columns show the full spectrum between 2.5 and 17.5 GHz, while the right-side column provides an amplification close to the peaks.

The addition of a metal-based layer follows similar overall trends to those observed in the ceramic-based case. Starting with the SMM filler, shown in Figure 5(a), we observe that it can enhance the previous MW absorption with the inclusion of thin (S) added layers. However, as the thickness increases, the absorption decreases. Additionally, a continuous frequency shift is observed as the thickness increases. On the other hand, the effect of adding an SMC is much more significant: absorption can be enhanced up to -35 dB (representing a 10 dB improvement) when the intermediate sheet (M) is used. As before, excessively thick layers degrade system performance. The frequency dependence of the peak position remains consistent with the layer thickness. As we did with the ceramic-based cases, we provide a clearer analysis of the effect of the addition of the secondary layer in Figure 6.

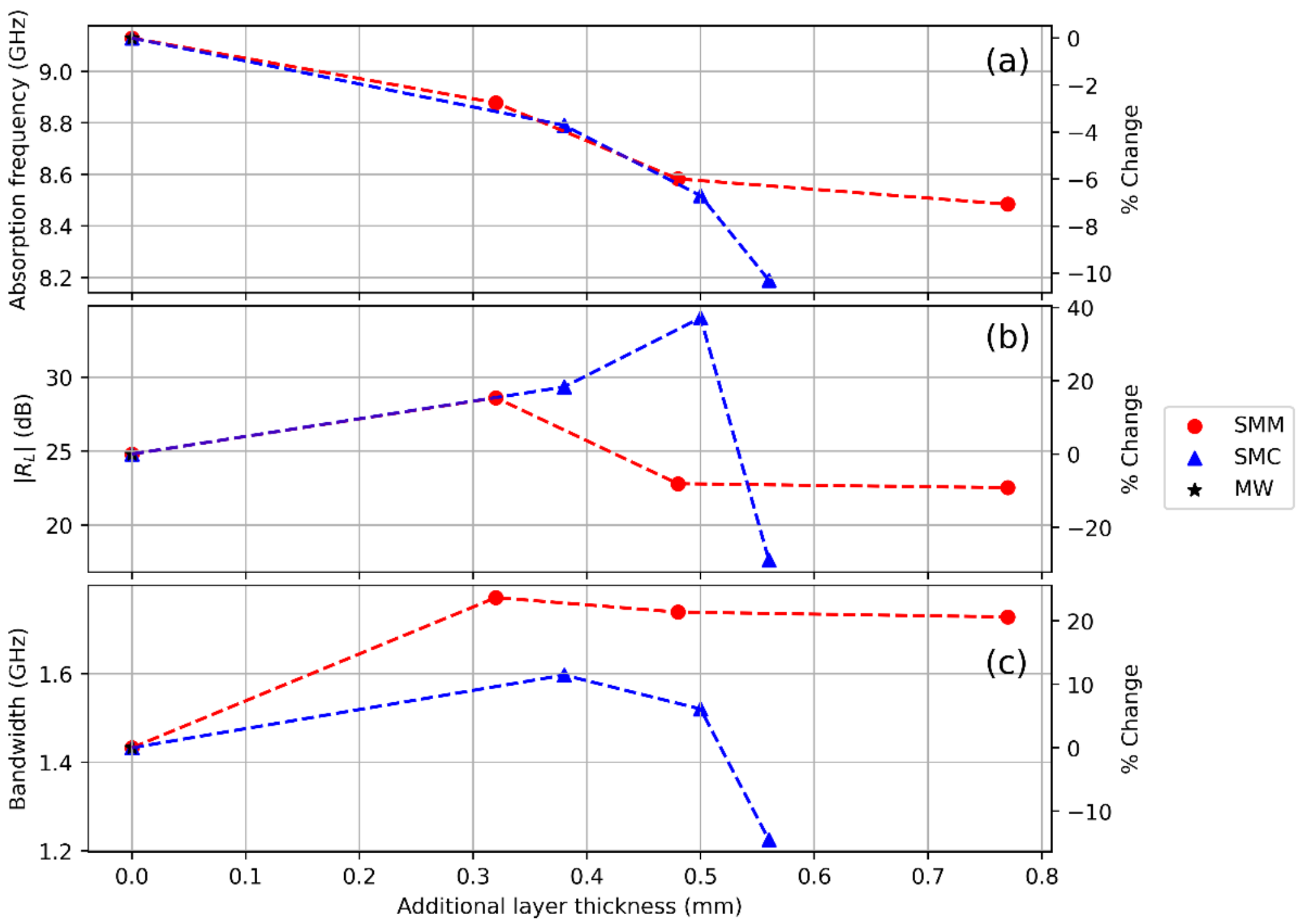

**Figure 6.** Dependence of the (a) absorption frequency, (b) maximum absorption, and (c) absorption bandwidth at -10 dB as a function of the thickness of the additional layer placed on top of the MW one (black star, at the origin) and its chemical composition, being SMM (red circles) or SMC (blue up triangles).

In comparison to the ceramic-based case, both materials now absorb at nearly the same frequency until the added layer exceeds 0.5 mm in thickness. Beyond this point, the effects of each material begin to diverge. Since the SMC consists of a central Fe core surrounded by a phosphate-based insulating shell, its magnetic permeability is expected to resemble that of the SMM. This explains why the electromagnetic differences for thin layers are imperceptible. Figure 6(a) illustrates how both types of layers shift the absorption peak identically with the S and M sheets. Differences become noticeable only when L layers are used. The absorption frequency can be shifted by up to 1 GHz with sub-millimetric secondary layers.

In terms of absolute absorption [Figure 6(b)], the SMC is a promising candidate for enhancing the radar absorption of the system through the creation of a bilayer structure. It can improve the system's shielding capacity by 40% (on the dB scale, from 25 to 35 dB) when a ~0.5 mm thick layer is incorporated. However, for layers thicker than 0.5 mm, the attenuation decreases rapidly. The pure SMM sample can increase the attenuation by nearly 20%, reaching approximately 30 dB with 0.3 mm thick sheets.

SMC is clearly the best option for increasing the absorbed power; however, as shown in Figure 6(c) , SMM is the optimal choice for expanding the absorption bandwidth. The peak width at -10 dBs is increased by 20% compared to the former MW absorption when a secondary SMM-based layer is incorporated. On the other hand, SMC (which had an exceptionally good response increasing the absorption intensity), is not capable of expanding the absorption bandwidth as much as SMM does. Thus, SMC is the best choice for situations when high $R_L$ is required, but SMM is a better option when broadband absorption is required.

Overall, the absolute changes observed with both ceramic-based and metallic-based additional layers are of the same order of magnitude. However, it is important to note that, although the metallic-based sheets achieved the greatest increase in absorbed power (+40%, up to 35 dB), the hexaferrite-based options provided much more consistent performance. SMM and SMC offer significant improvements within a narrow range of geometrical conditions, while hexaferrites perform well across a broader range of thicknesses. Furthermore, the addition of the secondary layer and the subsequent creation of bilayer systems simultaneously enhanced both the total absorbed power and the absorption bandwidth.

## 4. Conclusions

The results presented in this paper demonstrate the significant potential of magnetic bilayer systems to enhance radar absorption compared to traditional single-layer configurations. The experimental data support recent fundamental works that highlight the potential of these systems for stealth applications.

Micrometric magnetic bilayer systems emerge as a highly effective and robust solution for controlled enhancement of total absorption. The ability to chemically and structurally modify ceramic ferrites—directly influencing their functional behavior—provides an excellent means to adapt the electromagnetic response of the material. Random-anisotropy magnets, in the form of magnetic nanocomposites, have shown superior shielding capabilities compared to pure barium hexaferrite. Both reflection loss and absorption bandwidth can be increased by 20% compared to the single MW layer. Our experimental findings align with recent studies highlighting the potential of random-anisotropy magnets as thin (<1 mm) broadband microwave absorbers, making these materials ideal candidates for lightweight electromagnetic shielding applications.

The use of soft metal-based functional layers has provided more significant improvements than ceramic-based ones. However, their performance is highly sensitive to precise control of the layer thickness. SMC proves to be an exceptional option for enhancing absorbed power, improving $R_L$ by 40%, from -25 to -35 dB. On the other hand, pure SMM excels at widening the absorption peak, achieving -10 dB absorption over a 2 GHz range.

In summary, the introduction of a secondary magnetic layer has notably improved the electromagnetic absorption properties of a well-established single-layer absorbing material. We observed simultaneous improvements in both total absorbed power and absorption bandwidth for both ceramic and metal-based composite layers. It is also noteworthy that all improvements were achieved with micrometric, low-load coatings ($ff_W = 4\%$), which may be of particular interest for lightweight applications. Finally, the strong potential to control these properties should be emphasized. Our results show a clear dependence on the secondary layer thickness and the electromagnetic properties of the fillers, which are influenced by their chemical composition. A consistent and robust shift in absorption frequency was observed, alongside the non-linear relationship between $R_L$, sample thickness, and electromagnetic response.

## 5. Acknowledgements

Jaume Calvo-de la Rosa, Joan Manel Hernàndez and Javier Tejada thank the U.S. Air Force Office of Scientific Research (AFOSR) [grant number FA8655-22-1-7049] for their financial support. Joan Manel Hernàndez thanks the Departament de Recerca i Universitats de la Generalitat de Catalunya [grant number 2021SGR00328]. Jesús López-Sánchez acknowledges the financial support from grant RYC2022-035912-I funded by MCIU/AEI/10.13039/501100011033 and by the European Social Fund Plus (ESF+). Pilar Marín acknowledges the financial support from Spanish Ministry of Economic Affairs and Digital Transformation through the project PID2021-123112OB-C21-MICIIN, to the Comunity of Madrid NanomagCOSt project (S2018/NMT-4321). Finally, all the authors would also like to thank the enterprise AMES S.L. for providing the necessary materials.

## 6. Data Availability Statement

The data underlying this article will be shared on reasonable request to the corresponding author.